\documentclass{article}
\usepackage{arxiv}
\usepackage[utf8]{inputenc} 
\usepackage[T1]{fontenc}    
\usepackage{hyperref}       
\usepackage{url}            
\usepackage{booktabs}       
\usepackage{amsfonts}       
\usepackage{nicefrac}       
\usepackage{microtype}      
\usepackage{lipsum}		
\usepackage{graphicx}
\usepackage{natbib}
\usepackage{doi}
\usepackage{float}
\usepackage{amsmath} 
\usepackage[affil-it]{authblk}

\title{Embedding-based Multimodal Learning on Pan-Squamous Cell Carcinomas for Improved Survival Outcomes}

\author[1,2,*,$\dagger$]{Asim Waqas}
\author[1,2,*]{Aakash Tripathi}
\author[3]{Paul Stewart}
\author[2]{Mia Naeini}
\author[6]{Matthew B. Schabath}
\author[1,2,4,5]{Ghulam Rasool}

\affil[1]{Department of Machine Learning, Moffitt Cancer Center \& Research Institute.}
\affil[2]{Department of Electrical Engineering, University of South Florida.}
\affil[3]{Department of Biostatistics and Bioinformatics, Moffitt Cancer Center \& Research Institute.}
\affil[4]{Department of Neuro-Oncology, Moffitt Cancer Center \& Research Institute.}
\affil[5]{Department of Oncologic Sciences, University of South Florida.}
\affil[6]{Departments of Cancer Epidemiology and Thoracic Oncology, Moffitt Cancer Center \& Research Institute.}

\renewcommand{\shorttitle}{PARADIGM}

\begin{document}
\maketitle

\renewcommand\thefootnote{}
\footnotetext{\textsuperscript{*}Equal contribution.}
\footnotetext{\textsuperscript{$\dagger$}Correspondence: \href{mailto:aakash.tripathi@moffitt.org}{asim.waqas@moffitt.org}}

\begin{abstract}
Cancer clinics capture disease information at varying spatial (genetic to molecular to tissue to organ and beyond) and temporal scales (disease progression). The existing bioinformatic mechanisms do not encapsulate the complete spectrum of the heterogeneous nature of the data, especially under the challenge of missing data modalities. We propose an embeddings-based relational learning through Graph Neural Network (GNN) that can efficiently learn from multimodal, heterogeneous datasets for improved prediction of clinical outcomes. Our framework, PARADIGM, (1) generates embeddings from multi-resolution datasets using modality-specific foundation models, (2) aggregates the sample embeddings into patient-level representations in latent space, (3) fuses the aggregated embeddings into a unified graph representation, and (4) provides improved performance metrics for downstream tasks such as survival analysis. Our solution fuses unobserved but interrelated cancer variables in non-Euclidean space through a scalable framework. We train GNNs on survival prediction task using pan-Squamous Cell Carcinomas (SCC) in the head and neck (HNSC), lung (LUSC), bladder (BLCA), cervical (CESC), and esophageal (ESCA) cancers having 527; 496; 408; 294; and 184 patients, respectively. We also validate our approach for lung SCC data collected at Moffitt Cancer Center, comprising 103 patients. The data modalities for training and evaluation include EHR data (age at diagnosis, gender, ethnicity, race, smoking status, etc.), whole slide images, pathology reports, and molecular data (gene expression, miRNA expression, DNA methylation, DNA mutation, and protein expression). The model evaluation used the concordance index (C-index) of patient survival prediction and 7-fold cross-validation. We compared the multimodal GNN results with other unimodal and multimodal machine learning models, such as multilayer perceptron, Transformers, and XGBoost, and got improved predictions compared to other methods. We observe that the convergence of individual data modalities and integration across varying scales creates a unified view of the disease that is more insightful than the individual view or modality. Our solution aims to converge the entire spectrum of the disease and understand the patient’s genetic, physiological, and psychosocial circumstances in a unified framework. The proposed method can help the community by providing helpful insights on heterogeneous data integration and showing that the convergence of maximum data views across varying occurrences can accrue remarkable discoveries about the disease.
\end{abstract}

\keywords{Deep Learning \and Multimodal learning \and Oncology \and Graph Neural Networks \and Squamous Cell Carcinoma \and Embeddings}

\section{Introduction}
\label{sec:ch8:section1}

Clinical information about cancerous tumors is routinely recorded at different scales and resolutions throughout the progression of the disease, treatment, and survivorship \cite{chen2021multimodal, vanguri2022multimodal, boehm2021harnessing}. The resulting data may include multiple diverse modalities \cite{waqas2024multimodal}, including (1) molecular and -omics information recorded from genome, proteome, transcriptome, epigenome, and microbiome, (2) diagnostic radiological imaging, e.g., ultrasound, computed tomography (CT), magnetic resonance imaging (MRI), or positron emission tomography (PET) \cite{fass2008imaging-cancer}, (3) histopathology, immunohistochemistry (IHC), and immunofluorescence (IF) images and data, e.g., whole slide images (WSI) recorded from stained tumor tissue samples \cite{gutman2017digital, magi2018prostate-IHC}, and (4) Clinical data including Electronic Health Records (EHR) that consist of structured and unstructured data about the patient, their disease, clinical notes from routine visits, labs tests and vitals, and clinical reports from radiology, pathology or biopsy \cite{aschebrook2022overview-EHR, XIE2022-DeepLearning-EHR-SysReview}. Jointly learning from such multimodal, multiscale, heterogeneous information with the possibility of out-of-distribution (OOD) inputs (e.g., unknown disease or cancer type), incomplete, noisy, and missing data is challenging but crucial for tackling complex diseases such as cancer. The state-of-the-art multimodal artificial intelligence/machine learning (AI/ML) techniques use data fusion methods and various flavors of deep neural networks, including Transformers, convolutions neural networks (CNNs), multilayer perceptions (MLPs), etc. \cite{vanguri2022multimodal, chen2020pathomic, ahmed2023transformers, boehm2022multimodal-ovarian, luo2022multimodal-lung-cancer, chen2022pan, lipkova2022artificial, lu2022multimodal, xu2023multimodal, gao2020mgnn}. None of these models are intrinsically designed to handle heterogeneous multimodal datasets with noisy, incomplete, and missing data during training and after deployment. Given the enormous growth in the volume and variety of medical data and the inherent noise in the data recording procedures/instruments, there is a need to develop robust AI/ML models that can learn simultaneously from multimodal heterogeneous datasets to answer clinical questions.

\subsection{Multiscale, Heterogeneous Data in Cancer}
The challenge of understanding cancer exists at different scales, including (1) genetic and molecular aspects of cancer and its micro-environment, (2) pathological information about tissue, (3) radiological information about the organ, (4) physiology and health of the patient, and (5) their lifestyle. We must also understand the dynamic changes happening over time during the development and progression of the disease \cite{vanguri2022multimodal, acosta2022multimodal}. The advanced MedTech hardware/software allows us to take snapshots of cancer development from a normal cell to a pre-malignant lesion to a malignant tumor in many different ways and at different scales and resolutions. The challenge is to coherently ingest, process, denoise, and learn from these datasets \cite{AlNaqa2021, steyaert2023multimodal}. Integrating data from heterogeneous modalities is vital for creating a unified view of cancer, which can be more insightful and predictive than a view created by a single data modality. 

Cancer patients undergo various diagnostic imaging scans, lab tests, medical procedures, biopsies, and treatment regimens, including surgical resection, chemotherapy, radiation therapy, immunotherapy, or targeted therapies. During these activities, healthcare facilities collect and store various forms or modalities of data, which are later analyzed by medical professionals for treatment planning, disease monitoring, surveillance, and post-treatment survivorship. In this work, we focus on three types of data modalities: (1) digitized histopathology slides saved in the WSI file format, (3) -omics data that includes genomics, proteomics, and transcriptomics, (4) clinical data, which consists of patients' demographic information, clinical notes (including pathology reports), and lab results/vitals. Evidently, we are confronted with diverse data types that capture different yet complementary views of the underlying disease at different scales. That is, from molecular scale (captured using -omics data) to tissue (quantified with histopathology data) to demographic data (from EHR) and finally, a mixture of everything in semi-structured or unstructured text format (in the form of clinical notes and medical reports captured in the EHR). 

\subsection{Squamous Cell Carcinoma}
Squamous cell carcinoma (SCC) is a type of cancer that can arise in various organs and tissues beyond the skin, including the lungs, bladder, cervix, esophagus, and head and neck region. It arises from squamous cells, which are flat cells that line many surfaces in the body. Lung SCC accounts for approximately 15-20\% of all lung cancers, and head and neck SCC is the seventh most common cancer worldwide \cite{sabbula2023squamous, barsouk2023epidemiology}. These cancers can be aggressive and have significant mortality rates, highlighting the need for better understanding and treatment strategies. SCCs, particularly in the head and neck, can severely impact the quality of life due to their effects on essential functions like breathing, eating, and speaking. The survival rate for SCC is very high when detected early, with a 5-year survival rate of 99\% \cite{SCC-survival}. The management of advanced SCCs often involves a multidisciplinary approach combining surgery, radiation therapy, chemotherapy, and targeted or immunotherapies. Ongoing research aims to optimize these multimodal treatment strategies and identify the most effective combinations for different SCC types and stages. Researchers are actively exploring potential biomarkers for early detection, prognosis, and treatment response prediction in various SCCs \cite{campbell2018genomic, song2022proteomic}. In this work, we studied pan-squamous cell carcinoma (PanSCC) comprising lung, bladder, cervix, esophagus, and head and neck subtypes.

\begin{figure}[ht!]
     \centering
        \includegraphics[width=\textwidth]{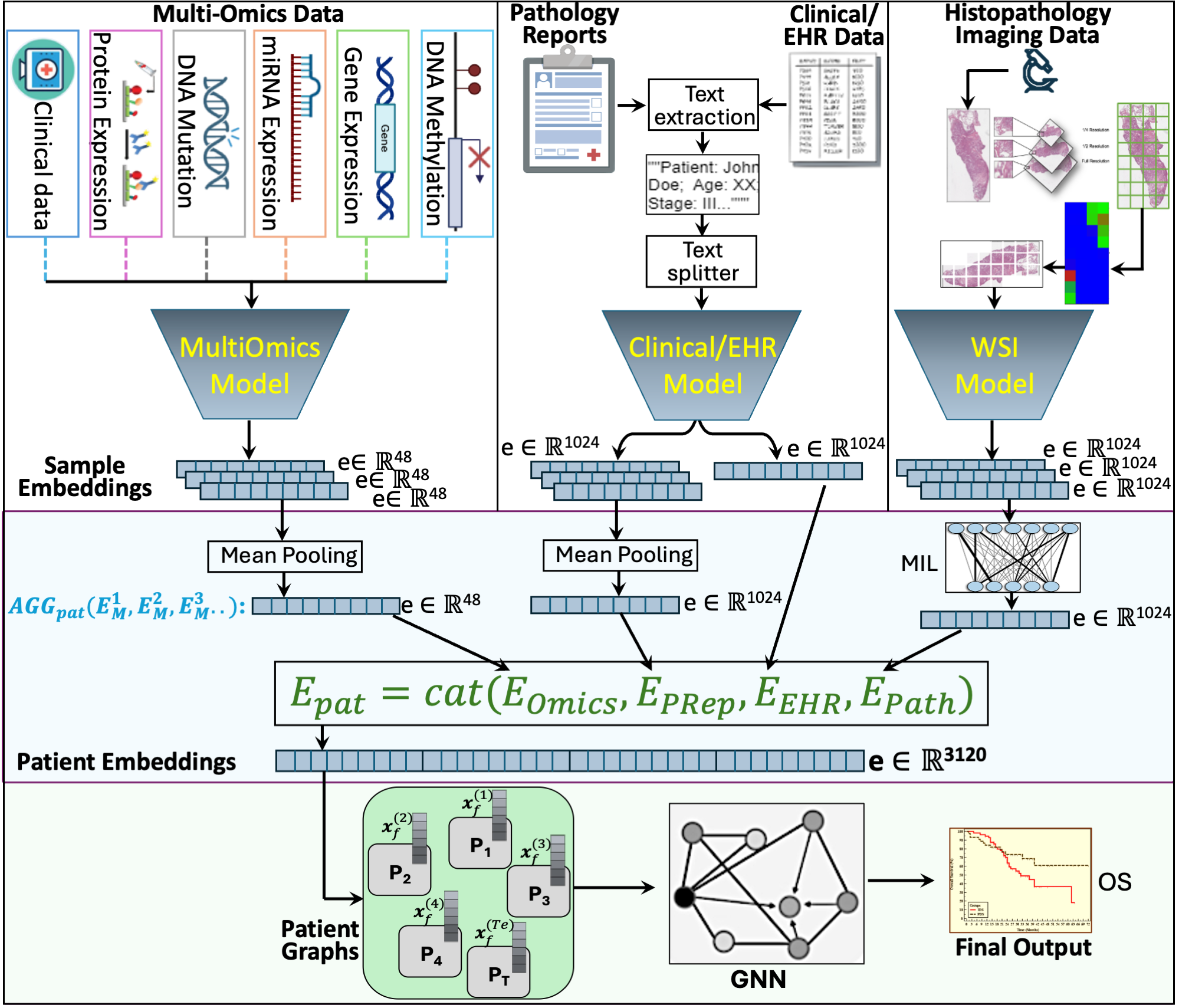}
        \caption{The schematic layout of the multi-modal data integration framework is presented. The framework integrates multi-omics data, pathology reports, clinical/EHR data, and histopathology imaging data to generate sample embeddings using modality-specific foundation models. These embeddings are pooled and aggregated in patient embeddings which are then used to construct patient graphs. These patient graphs are analyzed using Graph Neural Network (GNN) to predict overall survival outcome.} \label{fig:GNN-MML}
\end{figure}

\subsection{Multimodal Learning (MML)}
MML techniques combine information from various modalities to improve the accuracy and reliability of a given ML task \cite{acosta2022multimodal, kline2022multimodal, lipkova2022artificial, RadGenNets-2022, soenksen2022integrated, steyaert2023multimodal, pei2023review}. We can define five stages of multimodal learning, including preprocessing, feature extraction, data fusion, primary learning, and final classification \cite{MultimodalClassification}. The ``data fusion'' combines raw/extracted features or class prediction vectors from multiple modalities to create a single data representation. Data fusion can be performed in different ways: (1) \emph{early fusion} involves merging the modality-specific features (or embeddings) into a single feature vector before training AI/ML model, (2) \emph{intermediate fusion} involves training separate models for each data modality and combining model outputs for prediction, and (3) in \emph{late fusion}, the output of each model is used to produce a separate decision, which are then combined to make a final decision. The choice of fusion technique depends on the characteristics of the data and the specific problem being addressed \cite{acosta2022multimodal, jiang_sinha_aldape_hannenhalli_sahinalp_ruppin_2022, kline2022multimodal}.
\subsection{State-of-the-Art and Challenges in Oncology MML}
Modern and classical AI/ML models have been used to fuse various modalities of oncology data \cite{liu2022hybrid, hasanzadeh2022morel, vanguri2022multimodal,boehm2022multimodal-ovarian, luo2022multimodal-lung-cancer, chen2022pan, lipkova2022artificial, lu2022multimodal, xu2023multimodal, waqas2024multimodal, SeNMo}. However, these techniques are generally ad-hoc and limit their analysis to a selected set of modalities instead of using all available data. 

The state-of-the-art MML models use Transformer-based architectures to fuse and jointly learn from image and text data. Using these models out-of-the-box or their variants for fusing medical data that includes various types of radiology images, gigapixel histopathology/IHC/IF images, a variety of -omics data, and semi-structured EHR data is neither straightforward nor optimal.

Most MML models targeted for oncology applications are ad-hoc by design, use various types of AI/ML models and fusion methods subjectively, and involve a significant amount of manual feature/model engineering focused on a specific cancer type (or sub-type) and an organ \cite{liu2022hybrid, hasanzadeh2022morel, vanguri2022multimodal,  boehm2022multimodal-ovarian}. 

Oncology MML models generally target a limited number of data modalities. For example, radiology and genomics are fused for radio-genomics analysis, pathology and -omics are mixed and referred to as pathomics, or any one modality (e.g., radiology or pathology imaging) is fused with demographic data (e.g., age or smoking status) from the EHR  \cite{vanguri2022multimodal, li2022review-radiogenomics, brancato2022relationship, Zhang2022}. 

The current state-of-the-art models use varying methods to address the challenges of missing and noisy data, which is pervasive in medical datasets. These models are not designed and trained to be robust to routine changes in the data in medical settings, e.g., changes in the data recording protocols, machine hardware/software/firmware updates, changes in patients' demographics owing to changes in the algorithm deployment site or a new mutation or disease variant. 

Answering cancer-related clinical questions using multimodal oncology datasets has its unique challenges, nuances, and subtleties and therefore warrants its own unified data integration framework that can (1) handle data heterogeneity, missingness, and noise, (2) learn relations between different modalities of a patient and between patients, and (3) make accurate predictions.

\subsection{Contributions}
Our proposed framework has six distinct components. First, we introduce an embedding-based flexible and robust approach to tackle multi-modality cancer data. Second, we introduce graphs on the generated embeddings and perform graph structure learning to identify intra-/inter-cancer relationships/patterns using GNNs. Third, graph-based and learning-based methods with supervised techniques are introduced to fuse embeddings and handle missing/incomplete data. Fourth, the proposed graph structure enables multiscale learning across modalities. Fifth, we introduce a fusion mechanism that keeps the maximum amount of information intact while capturing relational patterns about the disease across different views of data. Sixth, we use the self-normalizing weights initialization on graph convolutional layers as well as use exponential linear unit (ELU) activation that ensures the self-normalizing property of GNNs. The evaluation metric for our framework is predicting accuracy for the given tasks of predicting OS quantified using the concordance index (C-index) \cite{uno2011c-C-Index}. Our work will potentially play a transformative role in the area of learning from multimodal, heterogeneous data in general and oncology data. The overview of our multimodal embeddings-based learning framework is illustrated in Figure \ref{fig:GNN-MML}.

\section{Methods}
We propose PARADIGM (\underline{P}an-Squamous Cell C\underline{a}rcinoma \underline{R}epresentation using \underline{A}dvanced Multimo\underline{d}al learn\underline{i}ng with \underline{G}raph-based \underline{M}odeling). Our proposed framework benefits from the state-of-the-art AI/ML models for learning modality-specific embeddings and combine these embeddings hierarchically using GNNs for learning inter- and intra-cancer patterns.

    \subsection{PARADIGM Architecture}
    We use standard mathematical notations, i.e., lowercase letters ($e$) for scalar values, lowercase bold ($\textbf{e}$) for column vectors, and uppercase letters ($E$) for matrices. The graphs are depicted by $G_\text{sub}$=$(V,C)$ having node-set $V$=$\{v_1, v_2,..., v_n\}$, where node $v$ has feature vector \textbf{x$_{v}$}, edge set $C$=$\{(v_i, v_j)\mid v_i, v_j\in V\}$, and subscript $_\text{sub}$ represents the purpose or task associated with the graph $G$. Fig. \ref{fig:GNN-MML} presents the main components of PARADIGM, (a) modality-specific sample embeddings, (b) aggregation and fusion into patient embeddings, (c) patient-specific graphs, and (d) joint graph learning. We describe these components in detail in the following:
    
        \subsubsection{Modality-Specific Embeddings}
        The first component of the PARADIGM architecture consists of a set of pre-trained, locally fine-tuned modality-specific AI/ML models, e.g., various types of Transformers, CNNs, SNNs, ViTs, etc. \cite{dosovitskiy2021an, Vaswani2017, yang2022large-GatorTron, SeNMo}. Our motivation is to leverage the state-of-the-art AI/ML pre-trained modality-specific models to find the most representative embeddings for all four data modalities. We prefer the models that have been pre-trained using related medical datasets \cite{yang2022large-GatorTron, azizi2022robustREMEDIS, UNI, waqas2023revolutionizing, wasserthal2023totalsegmentator, qiu2023large, wornow2023shaky}. After the first data processing step of modality-specific operations, we get: $E_{\text{Path}} = [\textbf{e}_{\text{p}}^{(1)}, \textbf{e}_{\text{p}}^{(2)}, \ldots, \textbf{e}_{\text{p}}^{(T_p)}]$, $E_{\text{Omics}} = [\textbf{e}_{\text{o}}^{(1)}, \textbf{e}_{\text{o}}^{(2)}, \ldots, \textbf{e}_{\text{o}}^{(T_o)}]$, $E_{\text{EHR}} = [\textbf{e}_{\text{e}}^{(1)}, \textbf{e}_{\text{e}}^{(2)}, \ldots, \textbf{e}_{\text{e}}^{(T_e)}]$, and $E_{\text{PRep}} = [\textbf{e}_{\text{pr}}^{(1)}, \textbf{e}_{\text{pr}}^{(2)}, \ldots, \textbf{e}_{\text{pr}}^{(T_{pr})}]$ for WSI patches, -omics, EHR, and pathology report datasets, respectively. We have $E_\text{Path} \in \mathbb{R}^{D_p \times T_p}$, $E_\text{Omics} \in \mathbb{R}^{D_o \times T_o}$, $E_\text{EHR} \in \mathbb{R}^{D_e \times T_e}$, and $E_\text{PRep} \in \mathbb{R}^{D_{pr} \times T_{pr}}$, which gives $\textbf{e}_{\text{p}} \in \mathbb{R}^{D_p}$, $\textbf{e}_{\text{o}} \in \mathbb{R}^{D_o}$, $\textbf{e}_{\text{e}} \in \mathbb{R}^{D_e}$, and $\textbf{e}_{\text{pr}} \in \mathbb{R}^{D_{pr}}$. Here, $T_p, T_o, T_e, \text{and} T_{pr}$ represent the total number of patients, and $D_p, D_o, D_e,$ and $D_{pr}$ represent the size of embedding vector for the WSI patches, -omics, EHR, and pathology reports datasets, respectively. $D_p, D_o, D_e,$ and $D_{pr}$ can have different values based on the complexity of the data modality. The EHR data is considered the baseline and is always available for all patients. Other data modalities may be missing, i.e., $T_p$, $T_o$, $T_{pr}$ $\le T_e$.
        
        \subsubsection{Concatenation and Aggregation}
        We used a two-step process (1) aggregate the sample embeddings into patient-level embeddings and (2) concatenate resulting embeddings across all modalities. In the first step, we aggregate the different samples' embedding vectors for the same patient into a single patient embedding for each modality. We use a learning-based aggregation function to combine all available WSI sample embeddings for each patient to produce $E_\text{M} = AGG_{\text{avail}} (E_\text{M}^{1}, E_\text{M}^{2}, E_\text{M}^{3}, ... , E_\text{M}^{u} )$, where $AGG_{\text{avail}}$ represents an aggregator function implemented with a multiple-instance learning network for modality $M$ having total samples $u$ for each patient. $AGG_{\text{avail}}$ is trained using the ground-truth data of 'age at index' and MSE loss function. For the -omics and pathology reports data modalities, we simply aggregate the sample embeddings by mean pooling to generate patient embeddings for the respective modalities. As the result of this aggregation stage, we get aggregated modality embeddings $\textbf{e}_{\text{o}} \in \mathbb{R}^{48}$, $\textbf{e}_{\text{p}} \in \mathbb{R}^{1024}$, $\textbf{e}_{\text{e}} \in \mathbb{R}^{1024}$, and $\textbf{e}_{\text{pr}} \in \mathbb{R}^{1024}$ for -omics, WSIs, EHR, and pathology reports, respectively. The second step involves concatenating the embedding vectors from different modalities into a single latent representation for each patient. The aggregated and concatenated patient embedding is represented by $E_\text{pat} = cat (E_\text{Omics}^{u}, E_\text{PRep}^{u}, E_\text{EHR}^{u}, E_\text{Path}^{u})$, where $cat$ represents the concatenation operation such that $ E_\text{pat} \in \mathbb{R}^{3120 \times T_e}$. Thus, the new embedding vector for each patient now has information from the initial modality embeddings, $E_\text{M}$.
        
        \subsubsection{Patient-Specific Graphs}
        We construct patient graph $G_F$ where each node of $G_F$ represents a patient and the node embeddings are given by $E_f = [\textbf{x}_f^{(1)}, \textbf{x}_f^{(2)}, \ldots, \textbf{x}_f^{(T_e)}]$. The (weighted) adjacency matrix for the patient graph is calculated using the Euclidean distance between embedding vectors \cite{stamile2021graph}. For the patient graph $G_F$, the adjacency matrix can be found using: $A_\text{pat} = Euc_{ij}(E_M) = \sqrt{\sum_{k=1}^{D_e}{\left(E_{M_{ki}} - E_{M_{kj}}\right)^2}}$. The patient graph has the number of nodes equal to the number of patients in the clinical (EHR) data $\lvert V_\text{pat} \rvert = T_e$. We also generated the adjacency matrix using cosine distance between patient embeddings for comparing the effect of selecting different distance metrics on the downstream analysis.
        
        \subsubsection{Joint Graph Learning}
        Finally, the joint patient graph is used to train GNN using supervised learning and the MSE loss function. As the result of the graph learning, the node feature vector for each patient is updated from its neighbors in the message passing mechanism.

\begin{figure}[ht]
     \centering
        \includegraphics[width=\textwidth]{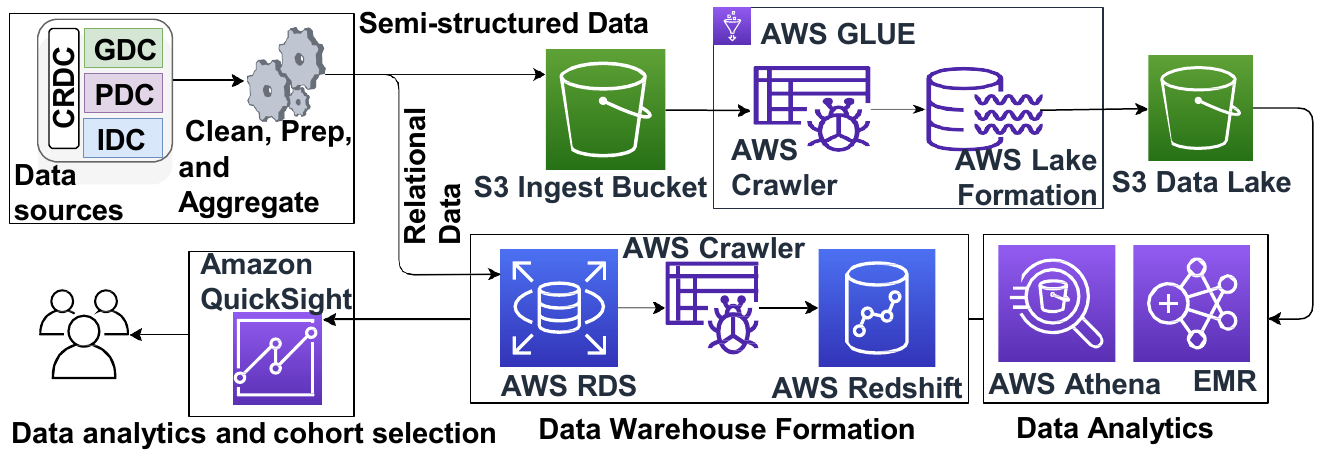}
        \caption{A schematic layout of the multimodal oncology database system (MINDS) \cite{MINDS} to collate, retrieve, harmonize, and serve multimodal data to PARADIGM models for training. We integrate data from National Cancer Institute (NCI)'s publicly available resources including, Cancer Research Data Commons (CRDC), Genomic Data Commons (GDC), Imaging Data Commons (IDC), and Proteomic Data Commons (PDC) \cite{CRDC, CPTAC, HCMI, TCGA}.} \label{fig:database}
    \end{figure}
    
    \subsection{Learning Strategy}
    Having presented the architecture of the PARADIGM framework, we now present the learning strategy of our framework as follows.
    
        \subsubsection{Modality-Specific learning}
        The modality-specific learning involves three tasks, (1) dataset development, (2) evaluating and selecting the modality-specific publicly available (foundation) models, and (3) transfer learning and fine-tuning of modality-specific models.

        \begin{enumerate}
            \item \textbf{Datasets Development}: Training multimodal models requires multimodal cancer datasets with EHR data as the base modality and one or more of the following: radiology, pathology, or -omics. We used public databases developed and shared by the National Institute of Health (NIH) and National Cancer Institute (NCI) and Moffitt's local datasets to train and validate our framework. Generally, the publicly available datasets may have only one or two data modalities. In case multiple modalities are present, they are scattered across different databases, and it is difficult to identify and link patients across databases. We use the already developed multimodal database, MINDS, to curate and build patient cohorts across five squamous cell carcinoma datasets (TCGA-LUSC, TCGA-HNSC, TCGA-CESC, TCGA-ESCA, and TCGA-BLCA) \cite{MINDS}. We acquired four different types of data modalities. The multi-omics data includes protein expression, DNA mutation, miRNA expression, gene expression, and DNA methylation, along with clinical data We used the preprocessing steps for multiomics data adopted by SeNMo \cite{SeNMo}. Pathology reports are processed through text extraction and text splitting techniques to obtain meaningful segments of text data. Clinical/EHR, which includes structured clinical records, and histopathology data comprising WSIs have been curated from the MINDS system \cite{MINDS}. We also evaluated the framework by training on the data collected at Moffitt ($T=103$) and consisted of EHR data (including age at diagnosis, gender, ethnicity, race, smoking status, year of diagnosis, vital status (alive/dead), and tumor cellularity), pathology images, and -omics data (included RNA-Seq expression and protein expression) \cite{paul2021multi}. 

            \item \textbf{Modality-Specific Models}: We leverage pre-trained models for initial modality-specific data processing. Based on our analysis, we have selected pre-trained models for obtaining initial embeddings after supervised and self-supervised fine-tuning of these models individually and jointly \cite{yang2022large-GatorTron, SeNMo}. For the WSIs, we evaluated various models and selected UNI for generating embeddings \cite{UNI}. For the clinical data such as EHR and pathology reports, we evaluated GatorTron and ClinicalT5 models \cite{yang2022large-GatorTron, ClinicalT5}. Based on our experiments, we selected GatorTron as our embedding model because of its superior performance on medical language tasks and text interpretation from pathology reports \cite{yang2022large-GatorTron}. For the multi-omics model, we selected the only model that has been trained on the five molecular data types across 33 cancer sites and is publicly available, SeNMo \cite{SeNMo}. All of these embeddings are publicly available through HoneyBee framework and hosted on Hugging Face \cite{tripathi2024honeybee}.

            \item \textbf{Fine-tuning Modality Models and Embeddings Generation}: The preprocessed data was then used to fine-tune the modality-specific models on different tasks such as OS, or self-supervised contrastive loss. After fine-tuning, we generated the embeddings for each modality model for each data sample. It is pertinent to mention here that these are sample-level vectors in the case of molecular and pathology report data and patch-level embeddings in the case of WSIs.             
        \end{enumerate}

        \subsubsection{Concatenation and Aggregation of Embeddings}
        As a result of the processing presented above, we get the initial embeddings for each data sample in each category of modality. Next, we combine these for each modality to build a single feature vector for each patient, referred to as patient-level embedding $E_M = AGG_{\text{avail}} (E_\text{M}^{1}, E_\text{M}^{2}, E_\text{M}^{3},..., E_\text{M}^{u} )$. In the case of patch-level embeddings for WSIs generated from the GatorTron model, we employed multiple-instance learning (MIL) where a neural network is trained in a weakly-supervised fashion to generate the slide-level embeddings \cite{MIL}. We experimented with attention-based MIL (ABMIL), mean-pooling, and max-pooling learning strategies and selected ABMIL for its superior performance in predicting the patient's age at diagnosis \cite{ABMIL}. The molecular and pathology reports embeddings had only a few patients with more than one sample, so we employed a simple pooling using the mean of the sample vectors. As the result of aggregation, we get the uniform-dimensional embedding vectors for each patient in each modality type. After aggregation, we concatenated the patient's modality embeddings into a single embedding vector per patient, $E_\text{pat} = cat (E_\text{Omics}^{u}, E_\text{PRep}^{u}, E_\text{EHR}^{u}, E_\text{Path}^{u})$. It is important to note here that the concatenation operation caters for the missing modalities by padding zeros to the missing feature vectors. This way, the framework takes the union of patients across different modalities to ensure that patients with missing data modalities, as in the real-world scenario, are not excluded from the analysis. The concatenation approach allows the retention of all the available information for the given patient. The embedding aggregation and concatenation operations are needed for three reasons, (1) to reduce the size of modality-specific embeddings to represent patient-level embeddings, (2) to combine embeddings from different data modalities in a joint space for pathology, -omics, and EHR data, and (3) to produce a joint embedding of the same size despite the possibility of missing some data modalities for some patients.

        \subsubsection{Patient Graphs and Joint Graph learning}
        The final set of embeddings are used as nodal features in the joint graph, which are trained to predict the selected endpoints of OS. It is important to highlight that the choice of the endpoint (i.e., OS) does not limit the PARADIGM framework. Based on the datasets, cancer type, and organ/site, our framework can support various other endpoints: tumor response rate, tumor shrinkage, disease-free survival, time to treatment discontinuation, toxicity, and time to next treatment. We use joint embedding for each patient to build a graph that represents one disease, i.e., one graph for each of the five cancer types. We then combine these subgraphs into a joint PanSCC graph data and use it to train the GNN. We used various measures to quantify the change in the embeddings, e.g., Euclidean or cosine distance. For the prediction of OS, we used convolution-based GNNs as they aggregate neighboring nodes’ embeddings through a stack of multiple layers \cite{xie2022selfsupervised-on-Graphs}. We tested commonly used spatial and spectral convolutional GNNs such as GraphSAGE \cite{hamilton2017inductive}, Graph Attention Network (GAT) \cite{velivckovic2017graph}, Graph Convolutional Network (GCN) \cite{kipf2016semi}, and Graph Isomorphism Network (GIN) \cite{xu2018powerful}. Based on our preliminary experiments, we selected the GCN model as our downstream model \cite{kipf2016semi}.            
                
            

\subsection{Model Evaluation}
We have five trained models based on our selection of five cancers (cervical, bladder, head and neck, lung, and esophageal) and one model for PanSCC. We also compare the performance of PARADIGM models with the single-modality, two-modality, and three-modality models to highlight the benefits of using the multimodal approach. We also compare the performance of our models with the state-of-the-art multimodal models. We evaluate the models' performance using the C-Index \cite{Cindex}, which compares the predicted outcome probabilities from the model with the ground-truth data. C-Index ranges from 0 to 1, where a value of 1 indicates the model can perfectly distinguish between individuals who experience an event earlier and those who experience it later. A value of 0.5 suggests that the model performs no better than random chance, while a value below 0.5 indicates that the model's predictions are inversely related to the observed outcomes. We use logrank statistical test to quantify the performance of the models to establish statistical significance.



 \begin{figure}[ht]
        \begin{center}
        \includegraphics[width=\textwidth]{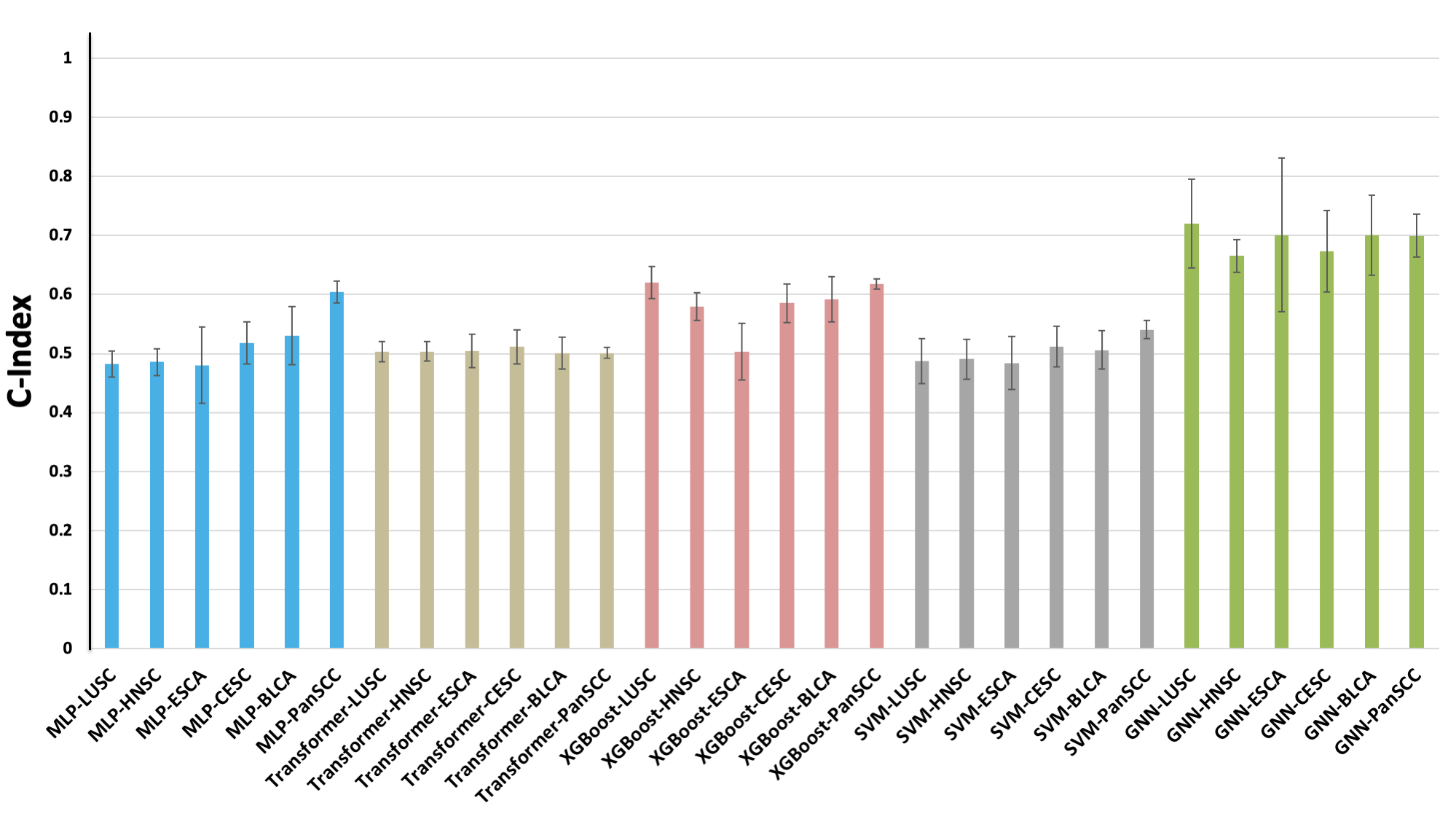}
        \caption{C-Index on training different models on the SCC data. The multimodal datasets consist of clinical, pathology reports, whole slide images (WSIs), and molecular data.}
        \label{fig:4Modal_Clin_Pathrep_Molec_WSI}
        \end{center}
    \end{figure}

\section{Results}

\subsection{Multimodal Integration of Oncology Database System (MINDS)}
Accessing, storing, pre-processing, harmonizing, clearing, and feeding multimodal data, consisting of various images, text, real numbers, and categories, to our models for training is a significant challenge. Given that no such system existed in oncology, we developed an in-house database system called MINDS and presented it in Fig. \ref{fig:database} \cite{MINDS}. MINDS can handle all types of data, including -omics (e.g., genomics, transcriptomics, and proteomics), diagnostic radiological imaging, histopathology/IHC/IF imaging, and EHR. MINDS employs data harmonization techniques to integrate all data types seamlessly and enables training PARADIGM models with up to petabytes of data. MINDS allows building cohorts for various cancer sites and types/sub-types. We followed the following scheme to build MINDS: 

\begin{enumerate}
    \item Data Ingestion: We collected (semi-)structured data from Genetic Data Commons (GDC), Imaging Data Commons (IDC), and Proteomic Data Commons (PDC) \cite{IDC, GDC, PDC}. AWS Glue was used to pre-process and aggregate data \cite{aws-glue, aws-crawler}.
    \item Data Storage: The pre-processed semi-structured data were uploaded to an Amazon S3 ingest buckets \cite{aws-s3}. The structured data was stored in an AWS RDS instance \cite{aws-rds}. 
    \item Data Lake Formation: We used AWS Data Lake Formation tool \cite{aws-lake-formation}, which automated transforming the semi-structured data stored in the S3 bucket into a query-able data lake using AWS Glue crawlers and AWS Athena \cite{aws-athena}.
    \item Data Warehouse Integration: The structured data stored in AWS RDS was integrated into a data warehouse using Amazon Redshift \cite{aws-redshift}.
    \item Cohort Selection and Data Export: We used AWS step functions to coordinate and manage the data export process for the selected cohort. We also provide manifest files to interact with GDC, IDC, and PDC and download all the unstructured data using their APIs.
\end{enumerate}

\subsection{Modality-Specific Models, Fine-tuning, and Transfer Learning}
We investigated various modality-specific models for our data modalities, pathology, -omics, pathology reports, and EHR. Our criterion for evaluating the representativeness or optimality of these modality-specific models includes their predictive performance against the ground-truth data of OS. As a result of preliminary analysis, we used: (1) UNI for histopathology images \cite{UNI}. UNI is a vision transformer model pre-trained on whole slide images using the DINOv2 self-supervised framework \cite{UNI}. We selected UNI for its ability to capture intricate details and patterns in histopathology images through self-supervised pre-training \cite{UNI}. (2) GatorTron for the EHR data (including clinical notes and pathology reports) \cite{yang2022large-GatorTron}. GatorTron is a large language model specifically designed for clinical NLP. It was trained on a massive corpus of 277 billion words, including 82 billion words from de-identified clinical notes and 195 billion words from various English texts. (3) Self-Normalizing Networks (SeNMo) for -omics data \cite{SeNMo}, which is a mini-foundation model that has been trained on five molecular data types for more than 13,000 patients across 33 cancer types. Our experiments showed that UNI and GatorTron worked well with fine-tuning using small datasets \cite{yang2022large-GatorTron, azizi2022robustREMEDIS, paul2021multi}; however, SeNMo always required transfer learning with larger datasets, potentially linked to the complexity and variability of the -omics datasets \cite{SeNMo, waqas2024multimodal}. 

  \begin{figure}[ht]
        \begin{center}
        \includegraphics[width=\textwidth]{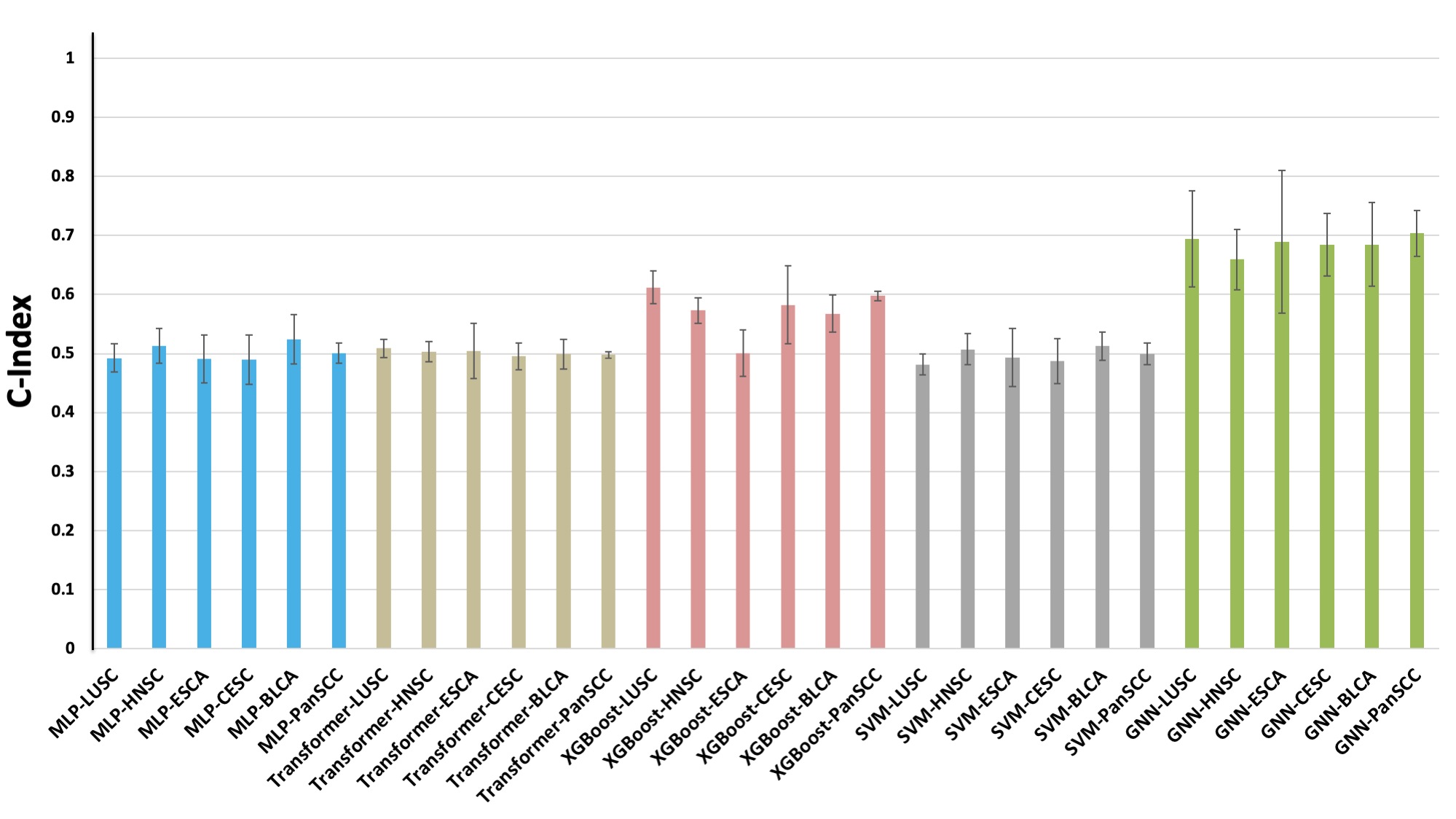}
        \caption{C-Index for OS predictions from different models on the SCC data comprising clinical, pathology reports, and WSIs data.}
        \label{fig:3Modal_Clin_Pathrep_WSI}
        \end{center}
    \end{figure}

    \begin{figure}[ht]
        \begin{center}
        \includegraphics[width=\textwidth]{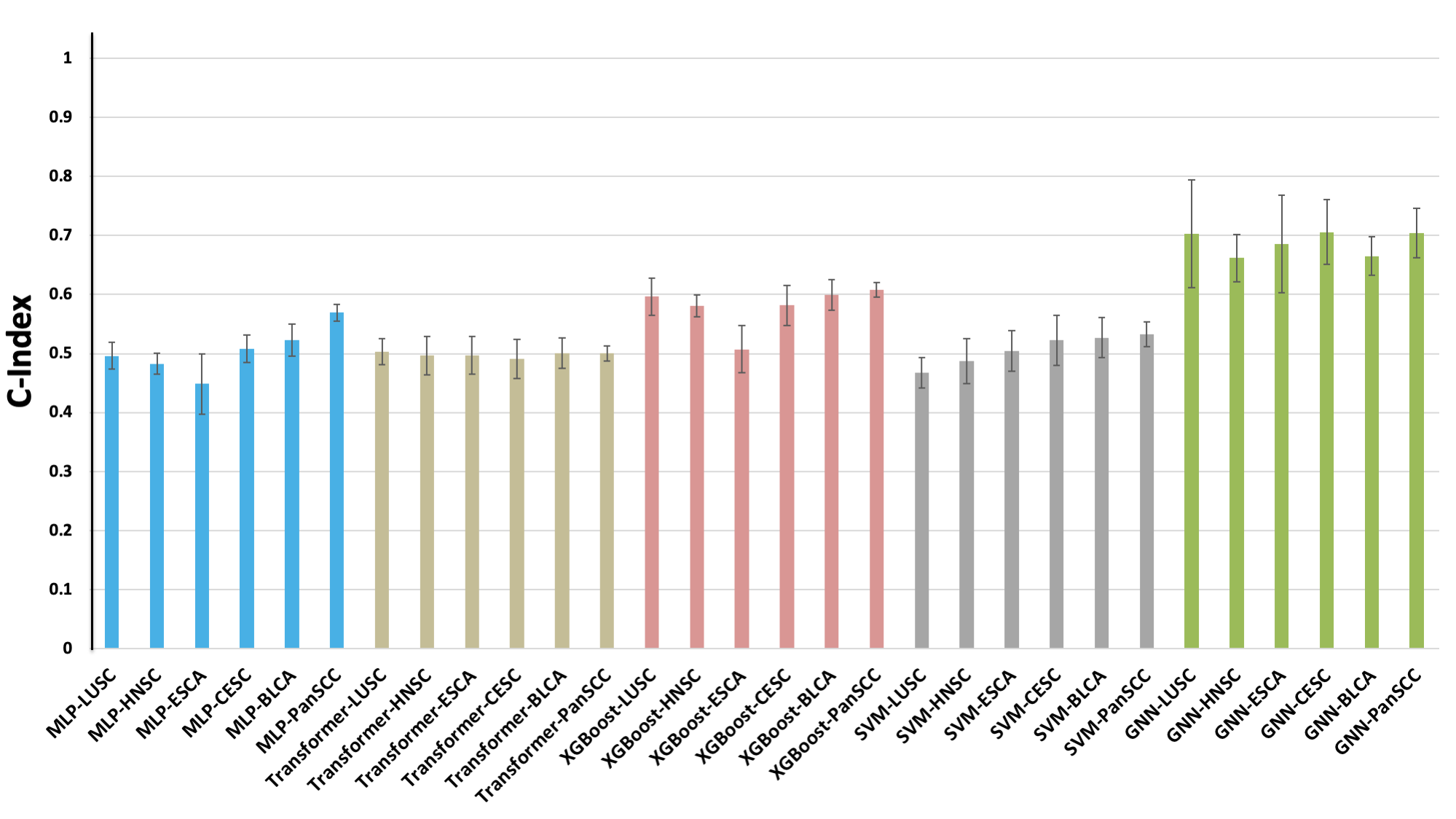}
        \caption{C-Index for OS predictions from different models on the SCC data comprising clinical, molecular, and WSIs.}
        \label{fig:3Modal_Clin_Mol_WSI}
        \end{center}
    \end{figure} 

    \begin{figure}[h]
        \begin{center}
        \includegraphics[width=\textwidth]{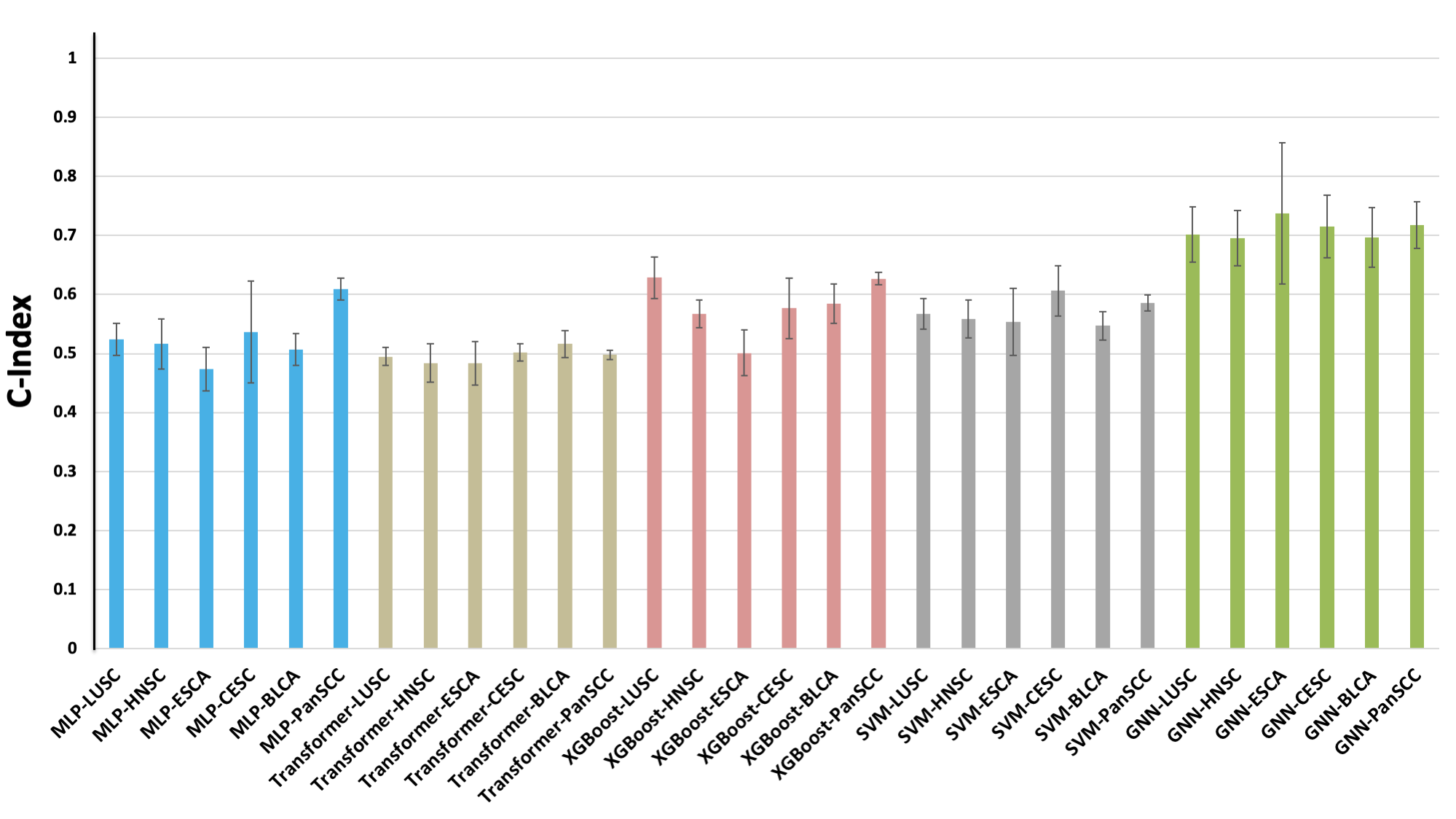}
        \caption{C-Index for OS predictions from different models on the SCC data comprising clinical, pathology reports, and molecular data.}
        \label{fig:3Modal_Clin_Pathrep_Mol}
        \end{center}
    \end{figure}

\subsection{Prediction of OS in PanSCC Cohort}
We predicted OS for pan-squamous cell carcinoma (PanSCC) patients using our framework, and a host of multimodal and uni-modal MLPs, Transformers, and classical ML algorithms such as eXtreme Gradient Boosting (XGBoost), and Support Vector Regression (SVR).

    \subsubsection{Four-modality Analysis}
    Figure \ref{fig:4Modal_Clin_Pathrep_Molec_WSI} shows the performance of different ML and DL models trained on multimodal datasets for SCC comprising four modalities; clinical, pathology reports, whole slide images (WSIs), and molecular data. The models include Multilayer Perceptrons (MLPs), Transformers, XGBoost, Support Vector Machines (SVMs), and GNNs. Each model was tested across various types of squamous cell carcinoma, lung squamous cell carcinoma (LUSC), head and neck squamous cell carcinoma (HNSC), esophageal carcinoma (ESCA), bladder carcinoma (BLCA), and pan-squamous cell carcinoma (PanSCC) comprising all five cancer types. The performance metric used is the C-Index for OS predictions, shown on the vertical axis. The MLP models, represented in blue, exhibit moderate performance with C-Index values around 0.5 to 0.6, with the best performance on the panSCC cohort. Transformer models, shown in beige, demonstrate similar performance to MLPs. The XGBoost models, depicted in red, generally outperform the MLP and Transformer models, achieving higher C-Index values, indicating better prediction accuracy. The SVM models, represented in gray, show consistent but slightly lower performance compared to XGBoost models. The GNN models, shown in green, achieve the highest C-Index values across all cancer types, suggesting that GNNs are particularly effective in leveraging multimodal data for accurate survival predictions. The error bars indicate the variability in the performance across different folds, providing insight into the reliability of the predictions. 

    \subsubsection{Three-modality Analysis}
    Figure \ref{fig:3Modal_Clin_Pathrep_WSI} illustrates the performance of different ML and DL models on multimodal datasets for SCC, encompassing clinical data, pathology reports, and WSIs. The MLP models, show moderate performance with C-Index values around 0.5, indicating the random predictions. Transformer models, depicted in beige, demonstrate similar performance to MLPs, with C-Index values clustering around the same range. The XGBoost models outperform both MLP and Transformer models, achieving higher C-Index values, which suggests better survival prediction accuracy. The SVM models, represented in gray, exhibit comparable performance to MLPs and Transformers. The GNN models achieve the highest C-Index values across all cancer types, significantly outperforming the other models. 

    Figure \ref{fig:3Modal_Clin_Mol_WSI} illustrates the performance of MLPs, Transformers, XGBoost, SVMs, and GNNs SCC data incorporating clinical data, molecular data, and WSIs. The MLP models show random performance for LUSC, HNSC, and ESCA with C-Index values $<$0.5, while a gradual improvement as we move from CESC, BLCA to panSCC data cohorts with C-Index $<$0.6. Transformer models do not perform well with all cohorts having C-Index $<$0.5. The XGBoost models consistently outperform the MLPs, Transformers, and SVM models. GNNs achieved the highest C-Index values across all cancer types, significantly outperforming the other models. 

    Figure \ref{fig:3Modal_Clin_Pathrep_Mol} shows the MLPs, Transformers, XGBoost, SVMs, and GNNs trained on SCC data incorporating clinical data, pathology reports, and molecular data. The MLP models performed better than all previous three-modality cohorts having C-Index $>$0.5 in five out of six datasets. Transformer models have similar ranges in C-Indices as before. The XGBoost models generally outperform the MLP and Transformer models, achieving higher C-Index values. The SVM models have shown better performance in this cohort with comparable performance to the XGBoost models. The GNN models achieved the highest C-Indices across all cancer types, significantly outperforming the other models.
    
    \subsubsection{Two-modality Analysis}
\begin{figure}[ht!]
        \begin{center}
        \includegraphics[width=\textwidth]{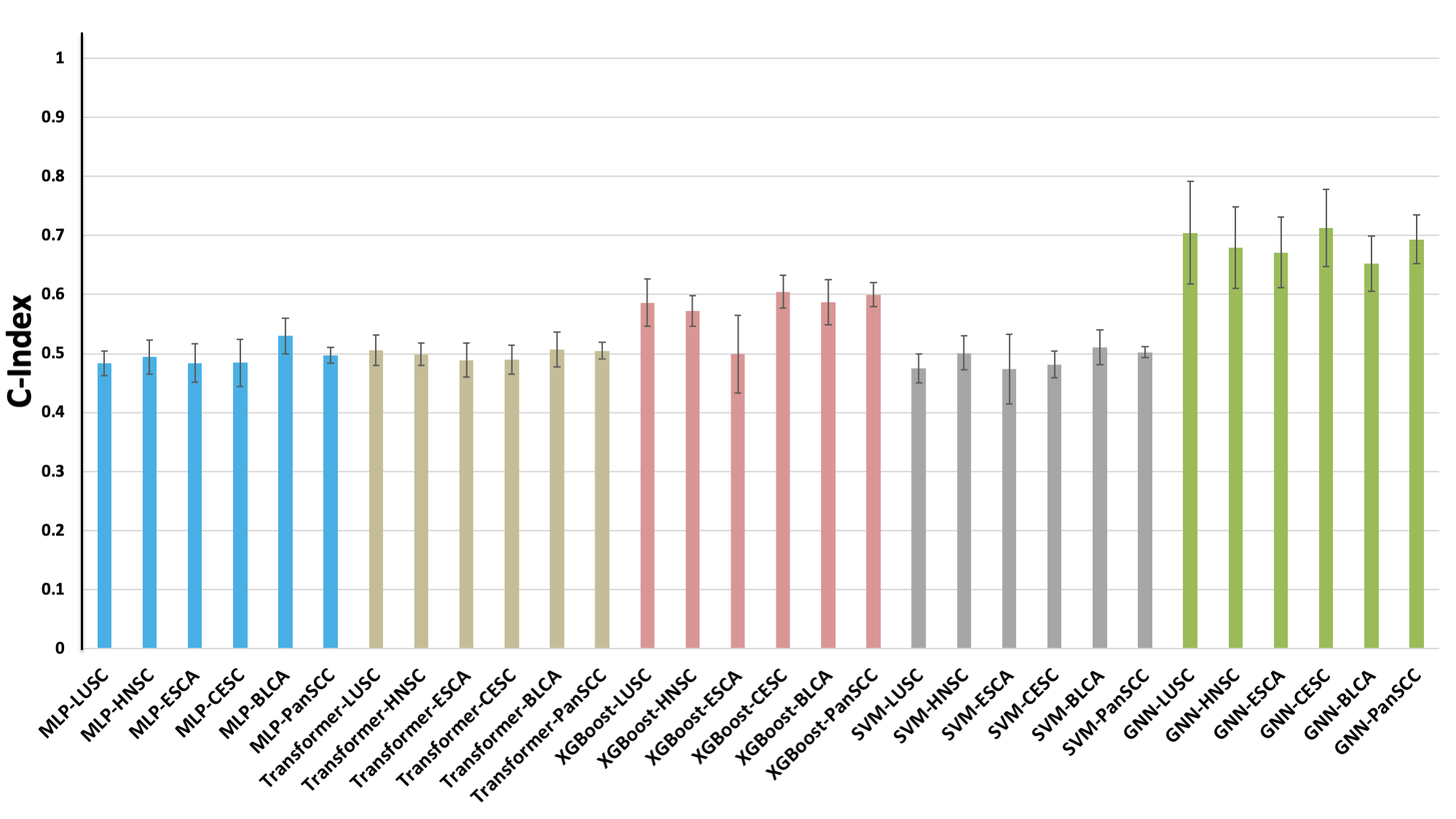}
        \caption{Performance Comparison of OS prediction by different models train on SCC datasets comprising clinical and WSIs data, measured by C-Index.}
        \label{fig:2Modal_Clin_WSI}
        \end{center}
    \end{figure}

    Figure \ref{fig:2Modal_Clin_WSI} shows the performance of MLPs, Transformers, XGBoost, SVMs, and GNNs applied to multimodal datasets for SCC, integrating clinical data and WSIs. The MLPs, Transformers, and SVM models exhibit random performance with C-Index $<$0.5 for most of the cohorts. The XGBoost models outperform the MLPs, Transformers, and SVMs, achieving C-Index values from 0.5 to 0.6. GNNs achieve the highest C-Index values across all cancer types, ranging from 0.65 to 0.72.
   
\begin{figure}[H]
        \begin{center}
        \includegraphics[width=\textwidth]{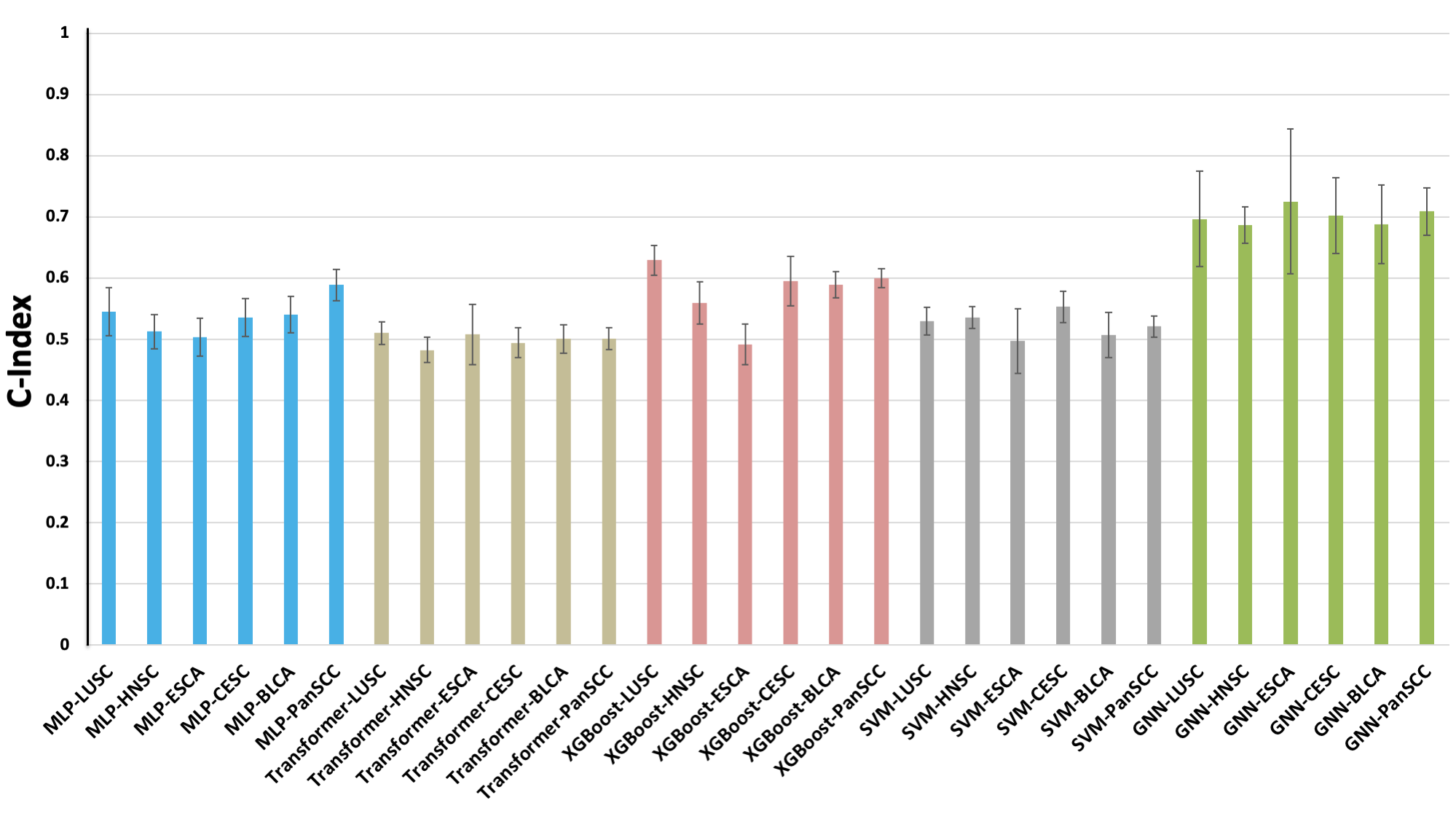}
        \caption{Performance Comparison of OS prediction by different models train on SCC datasets comprising clinical and pathology report data, measured by C-Index.}
        \label{fig:2Modal_Clin_Pathrep}
        \end{center}
    \end{figure}

    Figure \ref{fig:2Modal_Clin_Pathrep} illustrates that MLPs achieved the predictive performance for C-Index between 0.5 and 0.59, while XGBoost achieved values between 0.49 for ESCA to 0.63 for LUSC. Transformers fail to perform well, while SVMs have moderate performance range of 0.49-0.56. GNN models, illustrated in green, achieve the highest C-Index values across all cancer types, with all values around 0.7.

\begin{figure}[H]
        \begin{center}
        \includegraphics[width=\textwidth]{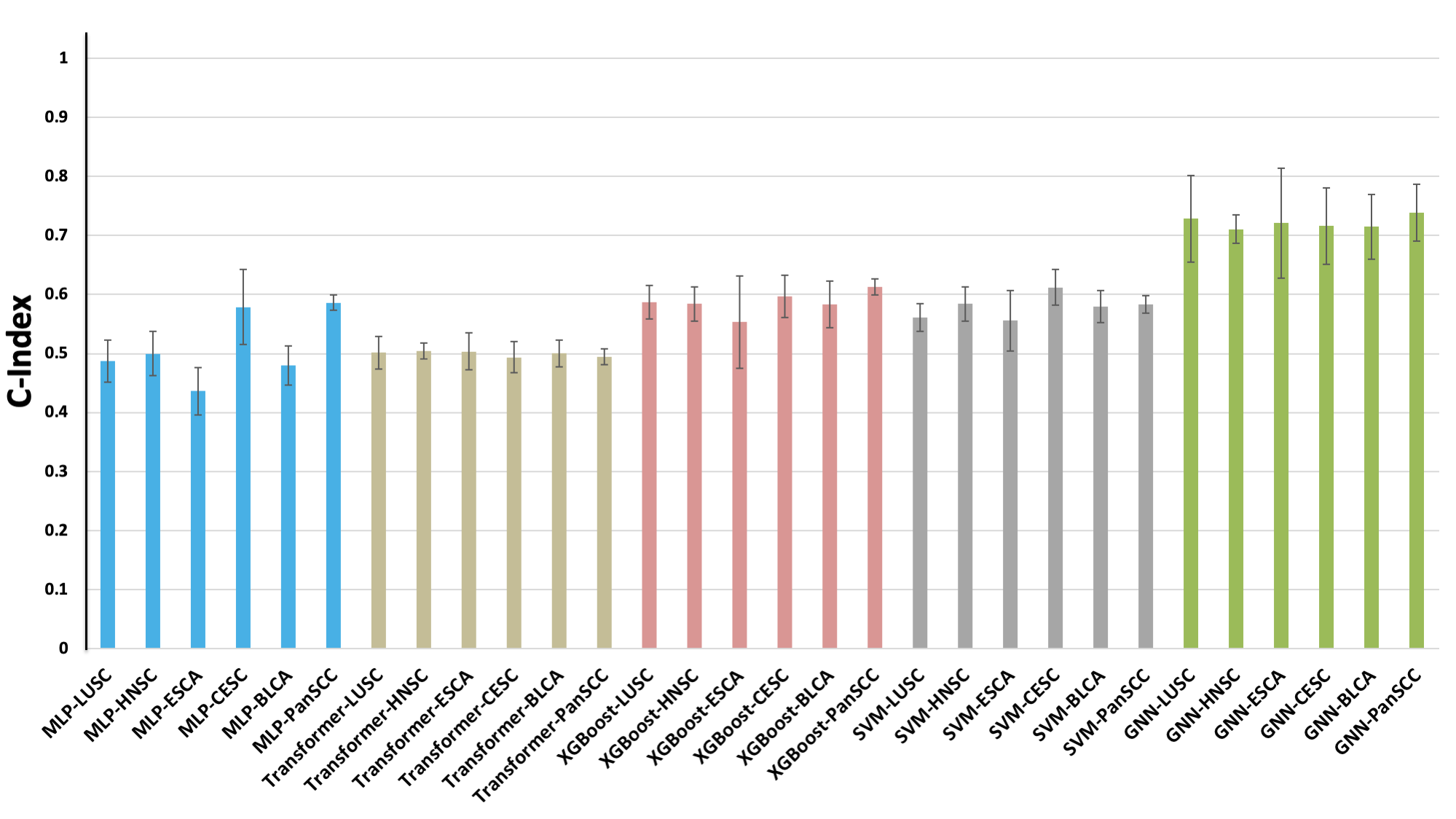}
        \caption{Performance Comparison of OS prediction by different models train on SCC datasets comprising clinical and molecular data, measured by C-Index.}
        \label{fig:2Modal_Clin_Mol}
        \end{center}
    \end{figure}

    Figure \ref{fig:2Modal_Clin_Mol} presents the performance of various models applied to SCC data with clinical and molecular data types. As with the previous results, the order of performance is Transformers, SVMs, MLPs, XGBoost, and GNNs, in increasing C-Index on predicting OS.

    \subsubsection{Internal Data Analysis}
\begin{figure}[h!]
        \begin{center}
        \includegraphics[scale=0.6]{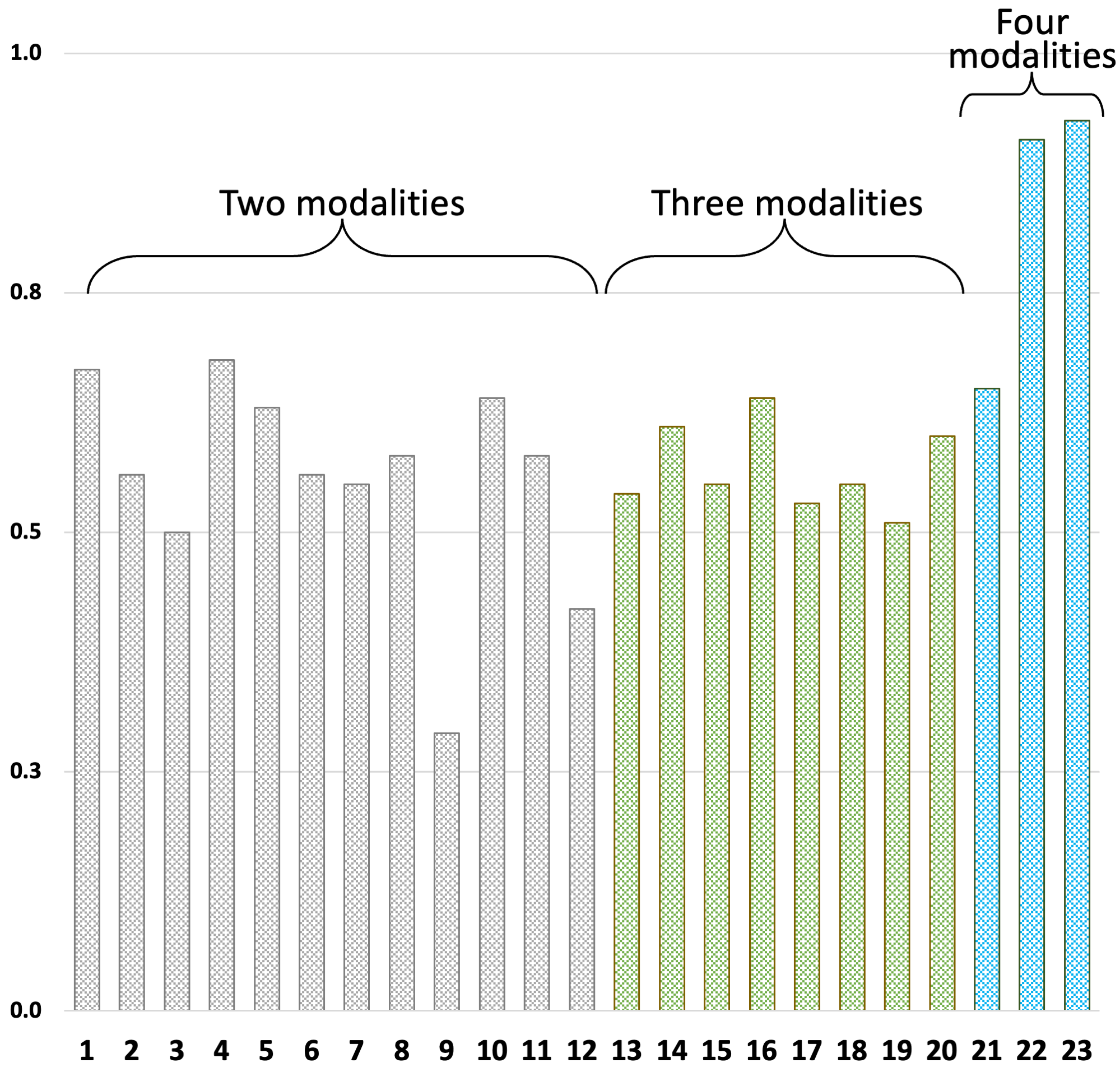}
        \caption{C-indices from different models trained combinations of modalities. The horizontal axis legend is a tuple: number-(model, modalities): 1-(MLP, EHR + RNA),  2-(MLP, EHR + Protein), 3-(MLP, EHR + Path), 4-(MLP, RNA + Protein), 5-(MLP, RNA + Path), 6-(MLP, Protein + Path), 7-(Transformer, EHR + RNA),  8-(Transformer, EHR + Protein), 9-(Transformer, EHR + Path), 10-(Transformer, RNA + Protein), 11-(Transformer, RNA + Path), 12-(Transformer, Protein + Path), 13-(MLP, EHR + RNA + Protein), 14-(MLP, EHR + RNA + Path), 15-(MLP, EHR + Protein + Path), 16-(MLP, RNA + Protein + Path), 17-(Transformer, EHR + RNA + Protein), 18-(Transformer, EHR + RNA + Path), 19-(Transformer, EHR + Protein + Path), 20-(Transformer, RNA + Protein + Path), 21-(MLP, EHR + RNA + Protein + Path), 22-(Transformer, EHR + RNA + Protein + Path), 23-(GNN, EHR + RNA + Protein + Path).}
        \label{fig:schematic_gnns}
        \end{center}
    \end{figure}

    Figure \ref{fig:schematic_gnns} shows the results from analyzing Lung SCC data collected at Moffitt Cancer Center \cite{stewart2019proteogenomic}. The data consists of four modalities, including pathology images, clinical data, and two subtypes of molecular data, RNA-Seq expression, and protein expressions \cite{stewart2019proteogenomic}. We generated pathology embeddings using REMEDIS \cite{azizi2022robustREMEDIS} and EHR embeddings using GatorTron \cite{yang2022large-GatorTron}, without fine-tuning the models. We trained SeNMo for -omics data to generate RNA-Seq expression and protein expression embeddings. Model evaluation was done using C-index and 10-fold cross-validation, as shown in Figure \ref{fig:schematic_gnns}. The models selected for evaluation are MLP, Transformer, self-normalizing network (SNN), and GNN. For two-modality combinations, MLP models show varying performance, with C-Index values ranging from approximately 0.5 to 0.7. Transformer models with two modalities exhibit poor performance compared to MLPs. For the three-modality combinations, the models generally show improved performance, especially the Transformers models exhibit better performance compared to two modalities. The four-modality combinations exhibit the highest performance. The MLP model trained on EHR, RNA, Protein, and WSIs achieves a C-Index of around 0.65, indicating strong predictive accuracy. The Transformer model with the same four modalities achieves a high C-Index, 0.91. The GNN model further improves the C-Index to 0.93, demonstrating the benefit of integrating all four data modalities.

\section{Discussion}
Disease-related information for oncology resides at varying scales and resolutions of data. Clinicians routinely fuse such diverse information mentally for decision-making, but there are limits to human processing capabilities. On the other hand, AI/ML models struggle when ingesting such heterogeneous, multiscale information for critical decision-making. Graphs and GNNs have been shown to perform well in contextual learning as well as represent the robust architectures of deep neural networks \cite{waqas2022exploring, waqas2024multimodal}. Jointly learning from such multimodal, multiscale, heterogeneous information with the possibility of incomplete and missing data modalities is challenging but crucial for understanding and tackling complex diseases like cancer. Our proposed framework for multimodal learning from multiscale, heterogeneous oncology datasets, has been trained and validated for accurately and robustly predicting overall survival (OS) for different tumor types (e.g., head and neck, lung, esophageal, cervical, and bladder cancers). We have also evaluated the framework on the Moffitt Cancer Center \& Research Institute's internal lung squamous cell carcinoma cohort comprising 103 patients. Our framework is built on pre-trained AI/ML models, Graph Neural Networks (GNNs), and self-surprised and supervised techniques to learn from multimodal, heterogeneous datasets. Our multimodal framework's potential applications go beyond cancer and healthcare settings, such as the deployment of AI/ML in many mission-critical application areas, including aviation safety, autonomous control of vehicles, automated decision-making in human-machine integrative environments, data fusion applications, and financial systems. Overall, the results indicate that models trained with more modalities tend to achieve higher C-Index values, reflecting better predictive accuracy for OS panSCC data. The highest performance is observed with the four-modality combinations, particularly for the GNNs-based PARADIGM model. All models and processed data have been made available via Hugging Face and GitHub.

\subsection{Embedding-based Approach}
The proposed framework introduced an embedding-based flexible and robust approach for multimodal learning. It benefits from the open-source models, pre-trained with histopathology, -omics, and EHR data \cite{azizi2022robustREMEDIS, yang2022large-GatorTron, SeNMo, UNI}, later fine-tuning these models to use them as the feature (or embeddings) extractors. This helped with rich feature extraction from all modalities at a significantly reduced computational cost. 

\subsection{Graph Structure and Contextual Learning}
Our framework introduced graph structure and GNNs on these embeddings to perform intra-/inter-cancer learning from heterogeneous datasets. The GNN-based approach enabled the representation of patient embeddings in the form of graph-structured data, where nodes represented patients and weighted edges between nodes represented inter-patient similarities. The process of updating the features of the current node based on its neighbors and the strength of the edges between these nodes helped in intra-/inter-cancer learning even though neighboring nodes may represent different cancers.

\begin{figure}[ht!]
    \begin{center}
    \includegraphics[width=\textwidth]{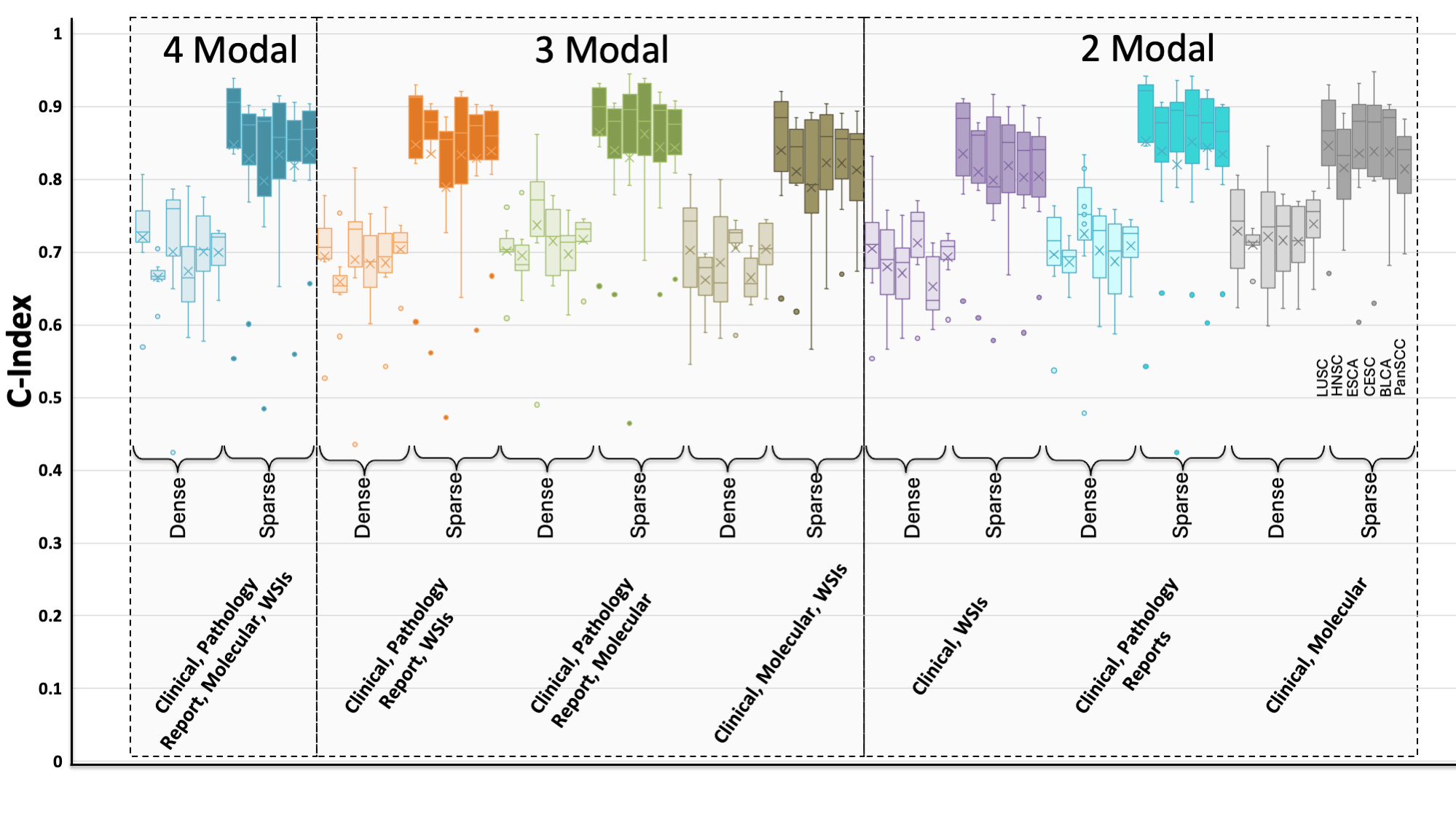}
    \caption{Impact of input graph sparsity on the predictive performance of GNNs. The C-Index is displayed for dense and sparse graph configurations using various combinations of clinical, pathology reports, molecular, and WSIs data.}
    \label{fig:graph_sparsity}
    \end{center}
\end{figure}

\subsection{Effect of Graph Sparsity}
In analyzing the graph structures and their effects on the model's performance, we varied the sparsity of the input graphs. Sparsity is controlled by thresholding the weights on the edges quantified by the distance metrics, such as Euclidean or cosine distance. Figure \ref{fig:graph_sparsity} illustrates the effect of changing the graph sparsity on the prediction of OS by the same GNN model. Across all combinations of modalities and cancer data types, the dense graph configuration generally achieves significant C-Index values. When we increase the graph sparsity, there is a notable increase in the model's performance, with C-Index values increasing to $>$0.8 in almost all the cases. This indicates that sparse graphs carry high signal-to-noise ratio and better predictive power for the GNN models, compared to dense graphs. Moreover, the inclusion of more data modalities generally enhances predictive accuracy, but the benefit is maximized when the graph is sparsely connected.

\subsection{Handling Missing Modalities}
The challenge of missing modalities is handled in our framework. Some patients may not have all the modalities; however, our framework allows learning with the available data and updating nodal features.

\subsection{Hierarchical Learning}
The framework's hierarchical nature enables it to handle sub-modalities seamlessly. We can add another data modality to fuse with the existing data types. For example, radiological scans such as MRI, CT, PET scans can be processed through REMEDIS foundation model to generate sample embeddings, or different stains of histopathology images such as IHC or IF slides can be added to the existing pipeline seamlessly. This innovative methodology of hierarchically fusing heterogeneous information based on the data modalities and sub-modalities is vital for learning from multiscale datasets that capture various aspects of underlying phenomena in space and time.

\section{Conclusion}
Our proposed framework, PARADIGM, demonstrates a significant advancement in the field of multimodal, heterogeneous data integration for cancer outcome prediction. By leveraging graph neural networks and modality-specific foundation models, we successfully generated comprehensive patient-level representations that capture the intricate relationships across various data modalities, including EHR, whole slide images, pathology reports, and molecular data. Our approach not only improved the predictive accuracy for survival analysis in pan-Squamous Cell Carcinomas (SCC) across multiple cancer types but also showcased the robustness and scalability of GNNs in handling complex, multimodal datasets. The superior performance of PARADIGM compared to traditional machine learning models underscores the importance of integrating diverse data sources to achieve a holistic understanding of cancer progression. Our findings suggest that the convergence of genetic, physiological, and psychosocial data into a unified framework can provide deeper insights and more accurate predictions, ultimately contributing to personalized and precision oncology. This research paves the way for future studies to explore the full potential of GNNs and multimodal data integration in various clinical applications, offering a promising direction for improving patient outcomes in cancer and beyond.

\bibliographystyle{unsrtnat}
\bibliography{references}  






\end{document}